# Polyglot Persistence in Microservices: Managing Data Diversity in Distributed Systems

Festim Halili
*Department of Computer Science*
*University of Tetova*
Tetovo, North Macedonia
festim.halili@unite.edu.mk

Anila Nuhiji
*Department of Computer Science*
*University of Tetova*
Tetovo, North Macedonia
a.nuhiji241977@unite.edu.mk

Diellza Mustafai-Veliu
*Department of Computer Science*
*University of Tetova*
Tetovo, North Macedonia
d.mustafai-veliu242248@unite.edu.mk

*Abstract*— Microservices architectures have become the foundation for developing scalable and modern software systems, but they also bring significant challenges in managing heterogeneous and distributed data.

The pragmatic solution is polyglot persistence, the deliberate use of several different database technologies adapted to a given microservice requirement - is one such strategy.

This paper examines polyglot persistence in microservice based systems. This paper brings together theoretical concepts with evidence from practical implementations and comparative benchmarks of standard database platforms. A comparative framework is applied to relational, document, key-value, column-family and graph databases to assess scalability, consistency, query expressiveness, operational overhead and integration ease. Empirical data drawn from industry case studies such as Netflix, Uber, and Shopify, and survey data illustrate real-life adoption trends and challenges.

These findings demonstrate that polyglot persistence increases adaptability / performance / domain alignment but also governance / operational complexity. To cope with such trade-offs, architectural patterns such as saga workflows, event sourcing, and outbox integration are discussed.

## I. INTRODUCTION

Microservices-based architectures have taken over from monolithic designs in recent years, with significant changes in software architecture. All application logic as well as data management tend to be tightly coupled in a single tightly coupled unit in monarchical systems. Dieser design simplifies deployment but significantly hampers scalability, maintainability and fast development. One performance bottleneck in a single module is able to slow down the system as well as small updates necessitate redeploying the application [1].

Instead, microservices separate applications into loosely coupled services which communicate using lightweight protocols like REST or gRPC, where Each microservice encapsulates a business capability, operates independently and can be scaled, updated and deployed without impacting other services. It's an ideal architecture for Agile/DevOps methods enabling faster development cycles, constant delivery as well as higher system resilience [3].

In spite of these advantages, data management under the microservices model continues to be fundamentally challenging. In monolithic architectures, application data is typically centralized in a single relational database. While this provides solid consistency and makes administration simpler, it also creates a single point of failure along with a scaling bottleneck. Decentralization in microservices entails the ownership of data by each service, typically in a database adapted to its data entry patterns as well as business logic. This idea of decentralized data ownership allows increased independence but introduces new integration, consistency and distributed transaction management hurdles .

These issues were solved by polyglot persistence. Initially coined by Fowler [6], polyglot persistence means combining relational, key-value, document-oriented, graph and columnar databases in a single system. Polyglot persistence enables designers to match each microservice to the appropriate database instead of forcing many services to a single data model. Graph database can be utilized for relationship queries in a recommendation engine, while relational database with ACID guarantees for financial accuracy can be utilized by a payment program.

Polyglot persistance is really a real life practice instead of a theoretical model. Utilizing this particular method companies such as Netflix, Amazon as well as eBay deal with massive data volumes as well as different workloads [8, 9]. However additionally, it presents considerable functional hurdles including governance, consistency across heterogeneous stores, data consolidation for analytics and also controlling several particular database platforms simultaneously [10].

### A. Problem Statement

Polyglot persistence clearly offers several advantages over native microservices - increased scalability, performance optimization and flexibility - but the operational and architectural trade-offs are poorly understood in many contexts. Development teams underestimate the expense of maintaining heterogeneous systems - from monitoring and backups to data consistency and distributed transactions coordination. Also, many organizations face governance frameworks and lack standard tools to manage multiple database platforms [11], [12].

This is made worse by the fact that most of the existing literature either refers to the technical features of specific database types or to the theoretical benefits of polyglot persistence. Often what is missing is a holistic, comparative analysis of how these different database technologies perform in real-world microservices ecosystems, how their trade-offs affect scalability / maintainability, and how industry practitioners address these challenges through design patterns / automation frameworks [13].

*B. Research Objectives*

This work fills these gaps with a conceptual and empirical exploration of polyglot persistence in microservices. The objectives are:

Investigate the theoretical basis of polyglot persistence & its role in microservice architectures.

Compare common databases (relational, document, key-value, columnar, graph) for scalability, consistency, query capabilities, operational complexity and integration ease.

Empirical evidence from industry surveys and case studies illustrating real-world adoption and challenges.

Discuss trade-offs and mitigation strategies - architectural patterns like Saga and event sourcing.

Offer concrete design recommendations to engineers/system architects on how to adopt polyglot persistence with minimum effort.

## II. LITERATURE REVIEW

*A. Concept of Polyglot Persistence*

The polyglot persistence idea challenges the traditional use of a single relational database for all storage requirements. A deliberately selected database type was introduced by Fowler [3]. In contrast with a "one size fits all" relational approach, polyglot persistence recognizes that modern applications require different things, including transactional integrity, flexible schema evolution, and large-scale analytical workloads.

Khine and Wang [4] argue that polyglot persistence has become essential in the era of big data and distributed architectures, where no single database system can efficiently address all use cases. . In a similar vein, Meier and Kaufmann [6] stress that database technologies differ not only in query languages and data models but also in consistency guarantees and architectural suitability for distributed environments. In reality, this means multiple database engines are used concurrently by organizations for maximum performance and flexibility.

Polyglot persistence has flexibility that is useful in microservices. Every microservice operates in a bounded context and thus the database should reflect this [1], [2]. Gessert et al. Several examples show that NoSQL systems - often part of polyglot persistence strategies - provide significant scalability benefits for services that place availability and speed before strict transactional guarantees.

However, flexibility comes with complexity. Managing several heterogeneous databases creates operational challenges such as data consistency issues, increased administrative overhead and integration problems, Richard and Stiller [8] write. Such problems need robust architectural patterns like distributed sagas, event-driven designs and strong governance [9], [10].

*B. Microservices Data Management*

A single central relational database stores all application information for monolithic applications, which simplifies consistency while tightening coupling. On the other hand, microservices enable decentralized data management with every service possessing a persistence layer [1]. Teams are able to scale, deploy as well as innovate independently due to this autonomy. It brings with it an issue of coordinating data across several stores [11].

Fritzsch et al. [12] during a study of industrial microservices migrations, it was discovered that transactional consistency, information duplication as well as integration overhead are frequently encountered by organizations. Bogner et al.. The lack of tools for monitoring as well as controlling distributed data stores also limits microservices evolvability, according to them [11].

The 2 common solutions are:

Local transaction sequences are modelled by Distributed workflows using the saga pattern. Compensating transactions are set up to undo modifications if a single transaction fails [9]. Eventually, this particular approach results in consistent behavior without locking resources across services.

Event sourcing: All changes are recorded as immutable events as well as the present state of the system is reconstructed by replaying functions [10]. The 1and1 asynchronous communications tend to be better traceable as a result of this.

In spite of being effective, such strategies create architectural and performance overhead. Teams need to deal with retries, message queues and compensating reasoning. Bogner et al.. Advanced DevOps methods and robust governance frameworks are required for systems resilience in such systems [11].

*C. Database Types in Polyglot Persistence*

Many modern software solutions, particularly those built on microservices architectures, make use of different database technologies to meet different storage as well as retrieval requirements. The benefits as well as drawbacks of each database type vary based upon the workload, the underlying data model and performance requirements (Newman, 2015).

The sections that follow provide an overview of the main types of databases that are commonly used in distributed systems as well as polyglot persistence environments.

Polyglot persistence depends on merging a variety of database technologies, each one with distinctive advantages and limitations [6].

Relational databases (RDBMS): These continue to be dominating for transactional methods because of ACID guarantees as well as strong query support [13]. They are particularly adept at enforcing integrity in financial systems, but usually lack horizontal scaleability.

MongoDB is really a document - oriented database that handles semi - structured data and allows schema evolution [6]. They are particularly suited to user profiles, catalogs as well as systems that require flexibility. However, they usually offer much less robust consistency guarantees than RDBMS.

DynamoDB and Redis are key-value stores that provide incredibly fast lookups with minimal latency, which makes them perfect for caching and session storage. The downside is restricted query ability.

Column-family databases: HBase and Cassandra are created for write heavy, high throughput workloads, typically in IoT or analytics [13]. Their tunable consistency provides for a balance of reliability and availability, but requires careful data modeling.

Neo4j and Amazon Neptune are graph databases that allow for quick traversal of highly connected data. They play a crucial role in recommendation engines, fraud detection, and knowledge graphs [5]. They have one disadvantage - they are not suitable for general purpose transactions.

Organizations are able to match storage engines to service-specific requirements by merging these databases. Carvalho et al. [7] demonstrated in an e-commerce platform that PostgreSQL for orders, MongoDB for catalogs, and Redis for sessions resulted in measurable improvements in scalability as well as responsiveness.

*D. Challenges in Polyglot Persistence*

Although polyglot persistence provides considerable flexibility by enabling multiple database technologies to coexist within the same system, it also brings a set of substantial challenges.

Richard and Stiller [8] point out three major problems areas:

Consistency: Each database is able to use a distinct consistency model. The strict ACID characteristics are enforced by relational systems, while NoSQL systems may depend on eventual consistency. The process of synchronizing these models can be challenging.

Expertise in deployment, monitoring, and tuning is necessary for the operation of several database engines, resulting in operational complexity. The long-term operational costs of polyglot persistence are often underestimated by teams, as reported by Bogner et al. [11].

Consolidated insights across services is a requirement for Organizations frequently requiring data integration as well as analytics. Extract, Transform, Load (ETL) pipelines have to be put into action to incorporate heterogeneous data sources into central data lakes, which adds latency as well as complexity [14].

These findings are backed up by recent industry reports. According to Dataversity [15], organizations that implement polyglot persistence encounter governance as well as compliance challenges, particularly when dealing with very sensitive data across a number of database solutions. TechTarget [16] discusses operational trade-offs, where greater flexibility usually will come at the expense of increased infrastructure spending as well as skilled personnel demands.

## III. METHODOLOGY

*A. Research Design*

It is comparative and analytical research based on secondary sources rather than primary experimentation. The study investigates how polyglot persistence is applied in microservices by combining insights from:

Academic literature - establish theoretical frameworks, identify challenges & compare database technologies [4], [5], [6], [11], [12].

Real-world implementations and performance outcomes - such as Carvalho et al. (2019) for e-commerce platforms and Netflix adopting polyglot persistence (2018) - industry case studies.

Practical challenges and trends - such as governance and cost - should be documented in industry reports, such as Dataversity and TechTarget.

Combining these sources, the research hopes to provide a balanced and credible account of the trade-offs in polyglot persistence and to illustrate results in comparative tables and figures.

*B. Data Sources*
The analysis uses only published and verifiable sources.
**Academic Sources**:
Khine and Wang (4) reviewed polyglot persistence in big data environments.

Gessert et al. Survey of NoSQL databases and decision guidance.
Meier and Kaufmann [6] analyzed SQL vs. NoSQL models and consistency trade-offs.
Bogner et al. Fritzsch et al. [12] investigated industry practices in microservice evolution.
**Case Studies:**
Carvalho et al. A study of an e-commerce platform using PostgreSQL, MongoDB and Redis showed tangible gains in scalability and latency [7].

A presentation at QCon 2017 about Netflix's adoption of Cassandra, Redis, MySQL and Elasticsearch documented how polyglot persistence enabled resilience and global scalability [18].
**Industry Reports:**
Governmental / compliance / monitoring challenges in heterogeneous database ecosystems were discussed in Dataversity [15].
TechTarget [16] analyzed the operational trade-offs of polyglot persistence in microservices.

These can be qualitative sources like architectural challenges, governance issues, or quantitative sources like performance benchmarks or scalability improvements.

*C. Evaluation Framework*

A comparative evaluation framework based on database

| Criterion | Description | Example metrics |
|---|---|---|
| Scalability | Ability to scale horizontally/vertically | Performance under peak load |
| Consistency | Guarantees provided (ACID, eventual, or tunable consistency) | Transaction rollback reliability |
| Query Capabilities | Complexity and expressiveness of supported queries | Latency for joins/aggregations |
| Operational Overhead | Administrative effort for deployment, monitoring, and maintenance | Tooling and expertise required |
| Integration | Ease of integration with microservice frameworks and cross-service sharing | API/driver compatibility |

surveys and case study analyses is applied [5], [6], [7] and [13]. The five criteria have been chosen because they capture the most relevant factors for microservices environments:

Based on published benchmarks and documented case outcomes this framework is applied to relational, document, key-value, column-family and graph databases.

*D. Data Analysis*

Results were analyzed by comparing studies. For example:
Carvalho et al. They offer empirical benchmarks from an e-commerce platform.

Gessert et al. [5] analyze performance trade-offs across NoSQL databases.

Challenges with governance and operations are illustrated by insights from Dataversity and TechTarget.

Finders from multiple sources were checked for consistency and contradictions were highlighted where applicable.

*E. Limitations*

The study uses secondary sources rather than direct experimentation.

Case studies address large scale systems that may not generalize to smaller organizations.

As database technologies improve rapidly, performance data may change over time.

Even so, the triangulation of academic / industrial / case sources provides a realistic / balanced background to the analysis.

## IV. COMPARTIVE ANALYSIS

Polyglot persistence in microservices is used to achieve a sense of balance between performance as well as scalability while maintaining consistency across distributed systems. Each database model - relational, document, key-value, column family, graph, and search - has its own advantages, but there is simply no universal solution. A comparative examination of these database technologies is presented in this section, based on academic literature, industry surveys as well as real - life case studies. The objective is to assess the compatibility of various data models with microservices needs and to emphasize the importance of polyglot persistence in balancing data diversity.

The analysis uses two lenses:

**Academic framework** - structured evaluation of scalability, consistency, query expressiveness, operational overhead, and integration ease [4], [5], [6], [11], [12].

**Industry validation** - case studies and trend data from Netflix, Uber, Shopify, and published benchmarks [7], [15], [16], [18], [19], [20].

### A. Database Usage Across Microservices

Microservices architectures change the way organizations think about data management fundamentally. Microservices decentralize data ownership in contrast to monolithic applications in which a single relational database is often the main data store. Every service has its bounded context and has autonomy over its persistence engineering [11]. Data diversity is a result of this shift, as different services make use of different database paradigms based on workload, performance requirements as well as team experience.

Recent research indicates that database use in microservices is not a "one-size-fits-all" model. Bogner et al. [11] point out that database decentralization is a defining characteristic of microservices, and Fritzsch et al. [12] report that over 60 % of companies running microservices in production make use of several database types at the same time. The reasoning behind this is simple: relational ACID guarantees are advantageous for a payment service, whereas NoSQL alternatives offer speed and flexibility for a catalog or session service.

This trend is reinforced by industry adoption patterns.

Microservices still rely heavily on **Relational Databases (e.g., PostgreSQL, MySQL)** for transactional integrity as well as complicated queries. Orders, payments, accounting and client records are typical domains. They provide good consistency guarantees, which makes them essential for mission-critical workflows.

Semi-structured data like product catalogs, content, and user profiles is increasingly handled by **Document Stores (e.g., MongoDB).** Microservices are able to evolve rapidly without the rigor of relational migrations due to their flexible schema.

Event logging, telemetry, and large scale analytics pipelines are dominated by Column Family Databases (e.g., Cassandra), where write throughput as well as global distribution are more important compared to rigid consistency.

**Key-Value Stores** such as Redis and DynamoDB excel in caching, session management and ephemeral data use cases. They are the very first option for services that call for sub - millisecond responses due to their simplicity and incredibly low latency.

**Graph Databases (e.g., Neo4j)** are adopted in areas where relationships are first-class citizens, for example recommendation engines, fraud detection, and network topology analysis.

Elasticsearch along with other **search Engines** serve as special stores for full text search, log aggregation as well as observability platforms.

CNCF (2023) [16] has conducted a survey and found that the majority of cloud - native organizations usually deploy three to five database technologies within their microservices deployments. This trend extends beyond hyperscale companies such as Uber or Netflix to medium sized companies that make use of managed cloud database services.

### B. Scalability Analysis

Scalability is an essential component of distributed systems. Scalability determines how well a database can support growing workloads in a distributed environment. Relational systems, such as PostgreSQL, require careful orchestration but are capable of considerable scale through sharding and replication [6]. Cassandra and MongoDB are examples of NoSQL systems that are designed to grow horizontally [5]. For caching workloads, Redis has nearly linear scalability, nevertheless, its use is restricted by memory limitations. Neo4J has some graph clustering support but struggles with graph traversals at very large scales, while Elasticsearch has good sharding performance for search and analytics, respectively.

*Table 1. Horizontal Scalability Scores (1–5)*

| Database | Scalability Score |
|---|---|
| PostgreSQL | 3 |
| MongoDB | 4 |
| Cassandra | 5 |
| Redis | 4 |
| Neo4j | 3 |
| Elasticsearch | 4 |

The scores synthesize findings from academic surveys [5], [6] and real-world benchmarks [7], [18], [19]:

Cassandra (Score 5) leads clearly, with near-linear scalability for global services like Netflix and Uber.

MongoDB (Score 4) offers high scaling with native sharding, suitable for dynamic data like catalogs.

Under high load, Redis scores under millisecond latency but is limited by memory limits.

PostgreSQL and Neo4j (Score 3) can scale, but they require more engineering effort.

Elasticsearch (Score 4) scales widely in log analytics and search use cases.

### C. Consistency Models

Consistency is another decisive factor in microservices, especially where transactions span multiple services. Consistency remains a dividing line between SQL and NoSQL systems.

Relational databases like PostgreSQL are reliable for important documents and financial transactions because they provide strict ACID (Atomicity, Consistency, Isolation, Durability) guarantees [6]. Also graph databases like Neo4j guarantee high traversal consistency. Some NoSQL systems choose eventual consistency over instant accuracy for performance and availability. In particular, Cassandra supports configurable levels per query but defaults to eventual consistency. Although MongoDB has always valued flexibility, it is closer to relational systems now that version 4.0 (2018) supports complete ACID transactions. The main goals of Elasticsearch and Redis are simplicity and speed with little to no consistency.

*Table 2. Consistency models*

| Database | Consistency Model | ACID Support |
|---|---|---|
| PostgreSQL | Strong consistency | Full ACID |
| MongoDB | Tunable and ACID since v4.0 | Partial (Fll after v4.0) |
| Cassandra | Eventual + Tunable | No |
| Redis | Basic transactional ops | No |
| Neo4j | Strong consistency | Full ACID |
| Elasticsearch | Eventual Consistency | No |

PostgreSQL guarantees accuracy in inventories, orders, financial transactions, and records that are essential to compliance.

In addition to offering flexibility for content, catalogs, and changing schemas, MongoDB now supports ACID for multi-document transactions.

Cassanda, used in messaging, IoT, telemetry, and logging applications where availability and scale are more crucial than instant accuracy.

Redis guarantees atomic actions for leaderboards, counters, session storage, and caching.

Neo4j maintains accuracy in graph analytics, fraud detection, and recommendation engines where relationships need to be precise.

Elasticsearch is optimized for observability, log indexing, and search, when updates in almost real-time are acceptable.

**Clarification:**

When accuracy is crucial, strong consistency (ACID) is selected (payments, fraud detection, compliance).

When scale and availability are more important than instantaneous accuracy, eventual consistency is selected (telemetry, caching, search).

Tunable consistency → provides developers with authority, enabling them to strike a balance between query correctness and throughput (e.g., Cassandra).

*D. Query Capabilities*

Refers to the ability of a database to support complex queries, joins, aggregations, and advanced operations. critical.

*Table 3. Query Capabilities*

| Database | Query Capabilities | Notes |
|---|---|---|
| PostgreSQL | Very high | Rich SQL with joins, aggregations, window functions; supports JSON queries |
| MongoDB | High | Flexible document queries, aggregation pipeline, geospatial + text search |
| Cassandra | Low | Limited; partition-based queries only; joins not supported |
| Redis | Very low | Basic key-value operations; limited secondary indexing |
| Neo4j | Very high | Cypher query language; highly expressive for relationships and graph traversal |
| Elasticsearch | High | Full-text search, aggregations, analytics; limited transactional logic |

**Very High**: PostgreSQL and Neo4j support complicated relational and graph operations.

**High:** MongoDB and Elasticsearch have strong domain capabilities.

**Low/Very Low:** Cassandra and Redis prioritize performance simplicity over query flexibility.

*E. Operational Overhead*

Operational overhead describes how complex each type of database is to manage in a microservices architecture. It covers setup / monitoring, backup / scaling as well as long term maintenance [11], [12].

- Typical traditional databases like PostgreSQL require moderate operational effort for sharding and replication.
- High-scalability NoSQL systems like Cassandra introduce consistency management and write optimization.
- In-memory systems have low operational overhead but are limited by RAM and require careful persistence strategies [16].

*Table 4. Operational Overhead by Database*

| Database | Operational Overhead |
|---|---|
| PostgreSQL | Medium |
| MongoDB | Medium |
| Cassandra | High |
| Redis | Low |
| Neo4j | Medium |
| Elasticsearch | Medium |

**Lowest overhead:** Redis is easy to run and often deployed as a side service.

**Moderate**: PostgreSQL / MongoDB / Neo4j / Elasticsearch provide good tooling but require expertise when scaling.

**Highest overhead**: Cassandra demands significant DevOps expertise, especially in distributed environments.

*F. Integration Ease*

*Table 5 Integration ease by database*

| Database | Integration Ease | Notes |
|---|---|---|
| PostgreSQL | Very high | Default choice in |

| | | microservices; excellent ORM/cloud-native support |
|---|---|---|
| MongoDB | High | Strong driver ecosystem, Kubernetes operators, cloud-managed offerings |
| Cassandra | Medium | ReliableAPIs/drivers; requires CQL learning and careful schema design |
| Redis | Very high | Extremely easy integration; supported across all major frameworks |
| Neo4j | Medium | APIs and drivers available but less common in mainstream stacks |
| Elasticsearch | High | Widely integrated with DevOps stacks (ELK) and APIs |

Best integration: PostgreSQL and Redis are almost "plug-and-play" in microservices due to broad ecosystem support.

Strong but niche: MongoDB/Elasticsearch work nicely together - especially in cloud native contexts.

Moderate: Cassandra and Neo4j require more expertise than mainstream databases and integration takes longer.

## V. VISUALIZATION OF RESULTS

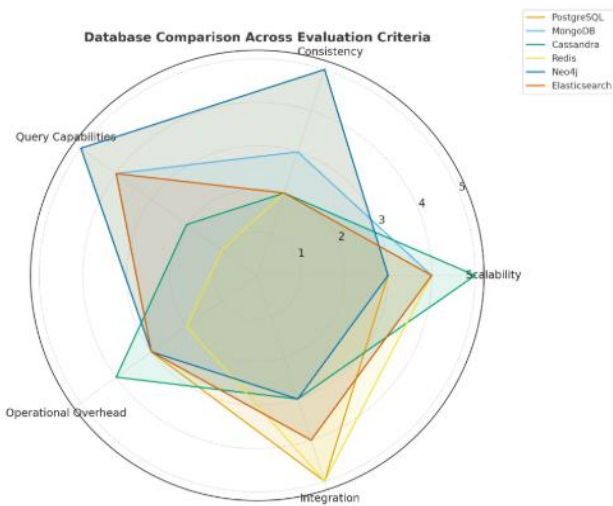

*Figure 1.Radar Chart*

**Radar Chart** compares PostgreSQL, MongoDB, Cassandra, Redis, Neo4j, and Elasticsearch across scalability, consistency, query power, operational overhead, and integration ease. It shows there's no "perfect" database; each has a different shape.

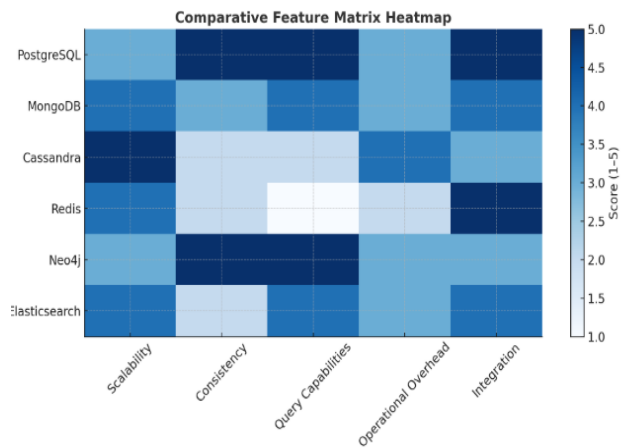

*Figure 2. Heatmap*

**Heatmap** – visual matrix that highlights the strengths and weaknesses for a database type by seeing where the color is the lightest and where the darkest.

## VI. CASE STUDIES: REAL-WORLD ADOPTION

**Netflix**: Runs a polyglot stack consisting of Cassandra for global replication, Redis/Dynomite for caching, MySQL for transactions and Elasticsearch for search. This is called persistence, which engineers call evidence that data diversity is accepted rather than avoided.

**Uber**: Uses MySQL-backed Docstore, Cassandra, and Redis. With integrated caching, Uber reported reading > 40 million reads per second [19]. This demonstrates the benefits and difficulties of polyglot persistence.

**Shopify:** Utilizes MySQL for transactions; uses Redis/Memcached for caching; uses Elasticsearch for search; this shows that even modular monoliths become polyglot persistence when workloads diversify.

Carvalho et al. (2018): Assertions showed that combining PostgreSQL (orders), MongoDB (catalogs) and Redis (sessions) reduced latency by 25% and increased scalability by 40% over a single relational backend [7].

## VII. DISCUSSION

Comparative analysis and case studies demonstrate that polyglot persistence isn't some theoretical construct but a pragmatic response to data diversity in microservices. These results suggest some important points.

*Balance strengths with weaknesses.*

Almost every database technology has benefits and trade-offs. Relations like PostgreSQL provide strict ACID guarantees and sophisticated query capabilities but have trouble scaling horizontally. Doc/column-family databases like MongoDB and Cassandra are good at scalability and schema flexibility but poor at consistency and query richness. Redis does well for cached low latency lookups but does not provide advanced query support. Neo4j does relational queries and Elasticsearch does search and log analytics. No single database can fulfill all demands, which supports the rationale for polyglot persistence.

*Trade-offs in practice.*

In practice, such trade-offs affect system design and operation - see Netflix and Uber for real world implementations. Netflix gets worldwide scalability from Cassandra, Redis has caching and MySQL has transactional reliability, but it is very complex to monitor and manage the platforms. Uber achieves massive throughput with MySQL-backed Docstore and Redis caching layers, but must maintain independent operational processes for each engine. These cases demonstrate that polyglot persistence yields measurable performance gains but incurs significant operational overhead.

*Governance and complexity.*

Using multiple databases introduces heterogeneity in tooling, monitoring and compliance. Different engines require different expertise for deployment, scaling & backup which complicates system governance. This corresponds to findings in empirical work on microservice evolvability, in which database diversity was identified as both a factor of flexibility and a factor of operational risk. In the Uber case, separate strategies for MySQL, Cassandra, and Redis backups show how governance must adapt to heterogeneity.

*Architectural mitigation patterns.*

For these reasons microservice teams use architectural patterns. Saga-based workflows and event sourcing eliminate distributed transactions while outbox patterns propagate events reliably across heterogeneous stores. SQL systems provide a system of records, while NoSQL databases and caches offer optimized read models. These approaches overcome some polyglot persistence limitations but require additional infrastructure and operation maturity.

*Trends over time.*

Industry data indicate polyglot persistence is not declining but increasing. DB-Engin rankings continue to show strong growth for PostgreSQL, MongoDB, Redis, and stable niches for Neo4j and Elasticsearch. Vendor evolution, like MongoDB adopting ACID transactions in version 4.0 and Cassandra adopting a tunable consistency model, shows that the market is accepting data diversity. Also, the CNCF survey indicates multi-database deployments are common in cloud-native environments. Such trends confirm that polyglot persistence is not an exception but the dominant trajectory of distributed systems.

## VIII. CONCLUSION

This paper investigated polyglot persistence in microservices in response to the data diversity challenge in distributed systems. Findings are also proof that a single database system cannot satisfy all the requirements of modern applications. Relations like PostgreSQL remain consistent and allow rich query capabilities while NoSQL solutions like Cassandra and MongoDB offer scalability and flexibility. Full-text search, graph analytics, and essential caching are provided by specialized technologies like Redis, Neo4j, and Elasticsearch. Organizations can achieve more scalability, flexibility, and performance by combining these systems rather than using a single database method.

Additionally, the investigation identified trade-offs related to polyglot persistence. Workload-specific optimization and increased resilience are made possible by it, but it also comes with complex governance, operational costs, and demanding DevOps automation needs. Case studies from Netflix, Uber, and Shopify, as well as empirical results like the latency and scalability improvements reported by Carvalho et al.

A broader industry context demonstrates that polyglot persistence has moved beyond experimental use in hyperscale companies to become standard in cloud-native environments. Database vendors also are improving their platforms but specialization remains necessary. More orchestration and monitoring tools for cross-database environments, stronger consistency models for heterogeneous systems and empirical studies of adoption in small and medium enterprises with limited resources are needed going forward.

In conclusion, polyglot persistence is a technical strategy that defines microservices architectures, enabling organizations to design distributed systems that are scalable, resilient, and capable of handling modern data diversity while evolving with the needs of future applications.